\documentstyle[prl,aps,epsfig]{revtex} \textheight24cm

\begin{document}

\draft
\twocolumn\wideabs{
\title{ Quantum diffusion of dipole-oriented indirect excitons 
in coupled quantum wells }

\author{A. L. Ivanov}
\address{ Department of Physics and Astronomy, Cardiff University,
Cardiff CF24 3YB, Wales, United Kingdom }

\date{May 06, 2002}
\maketitle
\begin{abstract}

A model for diffusion of statistically-degenerate excitons in (coupled) 
quantum wells is proposed and analysed. Within a microscopic approach, 
we derive a quantum diffusion equation, calculate and estimate the 
self-diffusion coefficient for excitons in quantum wells and derive a 
modified Einstein relation adapted to statistically-degenerated 
quasi-two-dimensional bosons. It is also shown that the dipole-dipole 
interaction of indirect excitons effectively screens long-range-correlated 
disorder in quantum wells. Numerical calculations are given for indirect 
excitons in GaAs/AlGaAs coupled quantum wells. 
\end{abstract}
\pacs{71.35.-y, 73.63.Hs}}
\narrowtext

\section{Introduction}
A system of indirect excitons in GaAs/AlGaAs coupled quantum wells (QWs) 
is a unique object for studing the transport and collective properties of 
interacting quasi-two-dimensional (quasi-2D) composite bosons. This is due 
to (i) a well-defined dipole-dipole repulsive interaction between indirect 
excitons, which is not sensitive to the internal spin structure of the 
particles (the exchange interaction is extremely weak so that the 
degeneracy factor $g=4$), (ii) the absence of quasi-2D excitonic molecules, 
(iii) strong suppression of the interface polariton effect and a long 
radiative livetime of excitons, and, as we show below, (iv) effective 
screening of in-plane QW disorder by interacting indirect excitons. 

The indirect excitons in coupled GaAs/AlGaAs have been extensively studied 
in various optical experiments (see, {\it e.g.},  
\cite{Butov1994,Butov2001,Butov2002,Negoita1999,Negoita2000,Larionov2001}). 
In particular, a specific jump of the signal intensity in photoluminescence 
(PL) dynamics of indirect excitons and its nonlinear behaviour 
have been interpreted in terms of Bose-Einstein (BE) stimulated scattering 
of statistically-degenerate Bose-particles \cite{Butov2001}. Furthermore, 
very recently a substantial progress has been achived in spatial confinement 
of quantum-degenerate indirect excitons by using a dip potential trap 
naturally-grown in GaAs/AlGaAs coupled QWs \cite{Butov2002}. A theory of the 
acoustic-phonon-assisted relaxation kinetics of statistically-degenerate 
QW excitons has been developed in \cite{Ivanov1999,Ivanov2000,Soroko2002}. 

For a dilute (quasi-) equilibrium gas of indirect excitons the degeneracy 
temperature $T_0$ and the chemical potential $\mu$ are given by 
\begin{equation}
\label{thermodynamic}
T_0 = \frac{\pi}{2} \left( \frac{\hbar^2}{M_{\rm x}} \right) n_{\rm 2d} 
\ \ \ \mbox{and} \ \ \ \mu = k_{\rm B} T \ln 
\left(1 - e^{-T_0/T} \right) \, , 
\end{equation}
where $M_{\rm x}$, $n_{\rm 2d}$ and $T$ are the in-plane translational mass, 
concentration and temperature of excitons, respectively. Crossover from 
classical to quantum, BE statistics occurs at $T \sim T_0$. The aim of the 
work is to develop a theoretical model for diffusive in-plane propagation of 
statistically-degenerate QW excitons. 

\section{Outline of the quantum diffusion theory} 
We assume that exciton-exciton interaction is dominant ($n_{\rm 2d} \geq 
10^9$\,cm$^{-2}$, according to the estimate given in \cite{Ivanov1999}) 
so that a system of quasi-2D excitons can be described in terms of the 
local quasi-equilibrium (thermodynamic) states. The self-diffusion flux 
density of QW excitons, ${\bf J}_{\rm x}^{\rm diff}$, is given by 
\begin{eqnarray}
\label{flux}
&&{\bf J}_{\rm x}^{\rm diff} = - D_{\rm x}^{\rm (2d)} \nabla n_{\rm 2d} - 
\frac{D_{\rm T}^{\rm (2d)}}{T} \nabla T \,, \ \ \mbox{where} \ \ 
\nonumber \\ 
&&D_{\rm T}^{\rm (2d)} =  \frac{T}{2} \left[n_{\rm 2d}  l_{\rm 2d} 
\frac{ \partial v_{\rm 2d}}{\partial T} - n_{\rm 2d} v_{\rm 2d} 
\frac{ \partial l_{\rm 2d}}{\partial T} \right] \, \ \ \mbox{and} 
\nonumber \\ 
&&D_{\rm x}^{\rm (2d)} = \frac{1}{2} \left[n_{\rm 2d}  l_{\rm 2d}  
\frac{ \partial v_{\rm 2d}}{\partial n_{\rm 2d}} +  l_{\rm 2d} v_{\rm 2d} 
- n_{\rm 2d} v_{\rm 2d} \frac{ \partial l_{\rm 2d}}{\partial n_{\rm 2d}} 
\right] \, . 
\end{eqnarray}
Here, $D_{\rm x}^{\rm (2d)}$ and $D_{\rm T}^{\rm (2d)}$ are the diffusion 
and thermal diffusion coefficients, respectively, $l_{\rm 2d}$ is the mean 
free path of excitons and $v_{\rm 2d}$ is their average thermal velocity. 

The mean free path of QW excitons is determined by $l_{\rm 2d} = 
v_{\rm 2d} \tau_{\rm x-x}^{\rm (2d)}$, where $\tau_{\rm x-x}^{\rm (2d)}$ 
is a characteristic scattering time due to exciton-exciton interaction. 
Thus one derives 
\begin{equation}
\label{length}
l_{\rm 2d} = v_{\rm 2d} \, \frac{2 \pi C \hbar^3}{|u_0|^2 n_{\rm 2d} 
M_{\rm x}} \, , 
\end{equation}
where $C$ is a numerical constant of the order of unity, $u_0/S$ is the 
potential of exciton-exciton interaction and $S$ is the QW area. 

For small transferred momenta in exciton-exciton scattering, $u_0$ is 
given by 
\begin{equation}
\label{potential}
u_0 = \pi \, \frac{\hbar^2}{\mu_{\rm x} \chi(d)} \, , 
\end{equation}
where $\mu_{\rm x}$ is the reduced exciton mass and $d$ is the effective 
separation between the electron and hole layers. The dimensionless 
function $\chi(d)$ has two well-defined limits, at $d \ll a^{\rm B}_{\rm 2d}$ 
and $d \geq a^{\rm B}_{\rm 2d}$ [$a^{\rm B}_{\rm 2d} = (\hbar^2 
\varepsilon_b)/(2 \mu_{\rm x})$ is the 2D exciton Bohr radius and 
$\varepsilon_b$ is the dielectric constant]. The first limit 
corresponds to the exciton-exciton exchange interaction in single QWs. 
In this case $\chi(d=0)$ is very close to unity \cite{Ivanov1999}, 
i.e., $\chi(0) = 1.036$, according to first-principle calculations 
\cite{BenTabou2001}. The second limit deals with a well-defined 
dipole-dipole interaction of indirect excitons in coupled QWs. In this 
case $\chi(d \geq a^{\rm B}_{\rm 2d}) = a^{\rm B}_{\rm 2d}/(2d)$ and 
Eq.\,(\ref{potential}) reduces to $u_0 = 4 \pi (e^2/\varepsilon_b) d$. 
The latter is consistent with the plate capacitor formula. Thus we 
approximate
\begin{equation}
\label{chi}
\chi(d) = \left\{
\begin{array}{ll}
1 \, , \ \ \ \ \ \  \ \ \ \ \ \ \ \ d \ll a^{\rm B}_{\rm 2d} \\
a^{\rm B}_{\rm 2d}/(2d) \, , \ \ \ \ \ d \geq a^{\rm B}_{\rm 2d} \, . \\ 
\end{array}
\right.
\end{equation}

Using Eqs.\,(\ref{length})-(\ref{potential}) one gets from 
Eq.\,(\ref{flux}):
\begin{eqnarray}
\label{diffusion}
&&D_{\rm x}^{\rm (2d)} = D_{\rm x}^{\rm (2d)}(T,T_0) = 
C \frac{\hbar}{M_{\rm x}} \chi^2(d) 
\left( \frac{\mu_{\rm x}}{M_{\rm x}} \right)^2 
\nonumber \\
&& \ \ \ \ \ \ \ \ \ \ \ \ \ \ \ \ \ \ \ \ \ \ \ \ \times \left[ 
\frac{1}{2k_{\rm B}} \frac{\partial E_{\rm kin}}{\partial T_0} 
+ 2  \frac{E_{\rm kin}}{k_{\rm B} T_0} \right] \, ,   
\end{eqnarray}
where the average thermal energy of QW excitons is given by 
\begin{equation}
\label{Ekin}
E_{\rm kin} = k_{\rm B} \frac{T^2}{T_0} \int_0^{\infty} 
\frac{z d z}{\exp[- \mu/(k_{\rm B}T) + z] - 1}
\end{equation}
with $\mu$ and $T_0$ defined by Eq.\,(\ref{thermodynamic}). 
In Fig.\,1 we plot $D_{\rm x}^{\rm (2d)} = 
D_{\rm x}^{\rm (2d)}(T,T_0)$ calculated for various $T$ and $T_0$ 
(note that $T_0$ is propotional to $n_{\rm 2d}$) by using 
Eqs.\,(\ref{diffusion})-(\ref{Ekin}) with $\chi = a^{\rm B}_{\rm 2d}/(2d)$ 
and $C=4/\pi$. In the limit of classical, Maxwell-Boltzmann statistics of 
QW excitons, when $T \gg T_0$, Eqs.\,(\ref{diffusion})-(\ref{Ekin}) reduce 
to 
\begin{equation}
\label{class}
D_{\rm x}^{\rm (2d,cl)} = 2 C \frac{\hbar}{M_{\rm x}} \chi^2(d)  
\left( \frac{\mu_{\rm x}}{M_{\rm x}} \right)^2 \frac{T}{T_0} \, . 
\end{equation}
Note that for a classical 3D gas of excitons one has $D_{\rm x}^{\rm (3d)}
\propto T^{1/2}$. In the limit of well-developed quantum statistics, 
$T \ll T_0$, Eqs.\,(\ref{diffusion})-(\ref{Ekin}) yield 
\begin{equation}
\label{quant}
D_{\rm x}^{\rm (2d,qm)} = \frac{\pi^2}{4} C \frac{\hbar}{M_{\rm x}} 
\chi^2(d) \left( \frac{\mu_{\rm x}}{M_{\rm x}} \right)^2 
\left(\frac{T}{T_0} \right)^2 \, . 
\end{equation}
Drastic reduction of $D_{\rm x}^{\rm (2d,qm)}$ in comparison with 
$D_{\rm x}^{\rm (2d,cl)}$ is due to non-classical accumulation of 
low-energy QW excitons which occurs at $T < T_0$. 

Substitution of Eq.\,(\ref{length}) into the expression for 
$D^{\rm (2d)}_{\rm T}$ [see Eq.\,(\ref{flux})] yields 
$D^{\rm (2d)}_{\rm T}=0$, because $l_{\rm 2d}(\partial v_{\rm 2d}/
\partial T) = v_{\rm 2d}(\partial l_{\rm 2d}/\partial T)$. This is in a 
sharp contrast with the 3D case, when $(\partial l_{\rm 3d}/\partial T) 
\simeq 0$ and $\partial v_{\rm 3d}/\partial T \propto T^{-1/2}$ give 
rise to the thermal diffusion coefficient $D^{\rm (3d)}_{\rm T} \propto 
T^{1/2}$ for Maxwell-Boltzmann statistics. Thus self-thermodiffusion of 
QW excitons is negligible in comparison with self-diffusion which is due 
to the concentration gradient. 

The drift flux density of QW excitons, ${\bf J}_{\rm x}^{\rm drift}$, 
is given by 
\begin{equation}
\label{drift}
{\bf J}_{\rm x}^{\rm drift} = - \mu^{\rm (2d)} n_{\rm 2d} 
\nabla \left( u_0 n_{\rm 2d} + U_{\rm QW} \right) \, , 
\end{equation}
where $\mu^{\rm (2d)}$ is the mobility of quasi-2D excitons and 
$U_{\rm QW}$ is the in-plane QW potential. The first term in the square 
brackets on the right hand side of Eq.\,(\ref{drift}) describes the drift 
flux due to repulsive exciton-exciton interaction. The exciton mobility 
is determined through $D_{\rm x}^{\rm (2d)}$ by
\begin{equation}
\label{mobility}
\mu^{\rm (2d)} = \frac{D_{\rm x}^{\rm (2d)}}{k_{\rm B} T_0} 
\left[ e^{T_0/T} - 1 \right] \, .
\end{equation}
For $T \gg T_0$ Eq.\,(\ref{mobility}) reduces to the Einstein relation, 
i.e., $\mu^{\rm (2d)} = D_{\rm x}^{\rm (2d)}/(k_{\rm B} T)$. In the 
quantum limit $T \leq T_0$, however, Eq.\,(\ref{mobility}) gives strong 
$n_{\rm 2d}$-dependent increase of the exciton mobility calculated in 
terms of $D_{\rm x}^{\rm (2d)}$. 

By using Eqs.\,(\ref{flux}), (\ref{drift}) and (\ref{mobility}), we end up 
with the nonlinear quantum diffusion equation:
\begin{eqnarray}
\label{final}
&&\frac{\partial n_{\rm 2d}}{\partial t} = \nabla \Big[ 
D_{\rm x}^{\rm (2d)} \nabla n_{\rm 2d} + \frac{2}{\pi} 
\left( \frac{M_{\rm x}}{\hbar^2} \right) D_{\rm x}^{\rm (2d)} 
\left( e^{T_0/T} - 1 \right) 
\nonumber\\ 
&& \ \ \ \ \ \ \ \ \ \ \ \ \ \ \ \ \ \times \ 
\nabla \left( u_0 n_{\rm 2d} 
+ U_{\rm QW} \right) \Big] - \frac{n_{\rm 2d}}{\tau_{\rm opt}} 
+ \Lambda \, , 
\end{eqnarray}
where the optical lifetime $\tau_{\rm opt} = \tau_{\rm opt}(T,T_0)$ of 
QW excitons is given by Eqs.\,(32)-(35) of \cite{Ivanov1999}, the 
self-diffusion coefficient $D_{\rm x}^{\rm (2d)} = 
D_{\rm x}^{\rm (2d)}(T,T_0)$ is determined by 
Eqs.\,(\ref{diffusion})-(\ref{Ekin}) and $\Lambda = 
\Lambda({\bf r}_{\|},t)$ is the generation rate of excitons. Note that 
the numerical constant $C \simeq 1$ in the expression (\ref{diffusion}) 
for $D_{\rm x}^{\rm (2d)}$ can be calculated by analysing the relevant 
linearized quantum kinetic equation with standard methods 
\cite{Lifshitz,Reichl1980} developed for classical bulk gases. 

\section{Screening of long-range-correlated disorder}

One of the most interesting results of the diffusion Eq.\,(\ref{final}) is 
effective screening of long-range-correlated QW disorder by dipole-dipole 
interaction of indirect excitons. In sharp contrast with excitons in single 
QWs, the blue-shift of the indirect exciton line, which has already been 
observed in many experiments (see, e.g., \cite{Butov1994,Negoita2000}), 
accurately fits the mean-field-theory result, $\delta U = u_0 n_{\rm 2d}$, 
over a broad range of concentrations, $10^{8}$\,cm$^{-2} \leq n_{\rm 2d} 
\leq 10^{11}$\,cm$^{-2}$. Here $u_0$ is given by Eq.~(\ref{potential}) with 
$\chi(d) = a_{\rm 2d}^{\rm B}/(2d)$.

In order to demonstrate and estimate analytically the screening effect, we 
put $1/\tau_{\rm opt} = 0$ (no optical decay) and $\Lambda = 0$ (no source of 
indirect excitons). In this case a steady-state solution of Eq.\,(\ref{final}) 
can be found analytically. The input, unscreened in-plane random potential 
$U_{\rm QW} = U_{\rm rand}({\bf r}_{\|})$ is due to the QW thickness and 
alloy fluctuations, ${\bf r}_{\|}$ is the in-plane coordinate. For average 
concentrations of indirect QW excitons such that $u_0 n^{\rm (0)}_{\rm 2d} 
\gg \ |U_{\rm rand}({\bf r}_{\|})|$ \ the \ steady-state \ solution \ yields

\begin{figure}
\centerline{\psfig{file=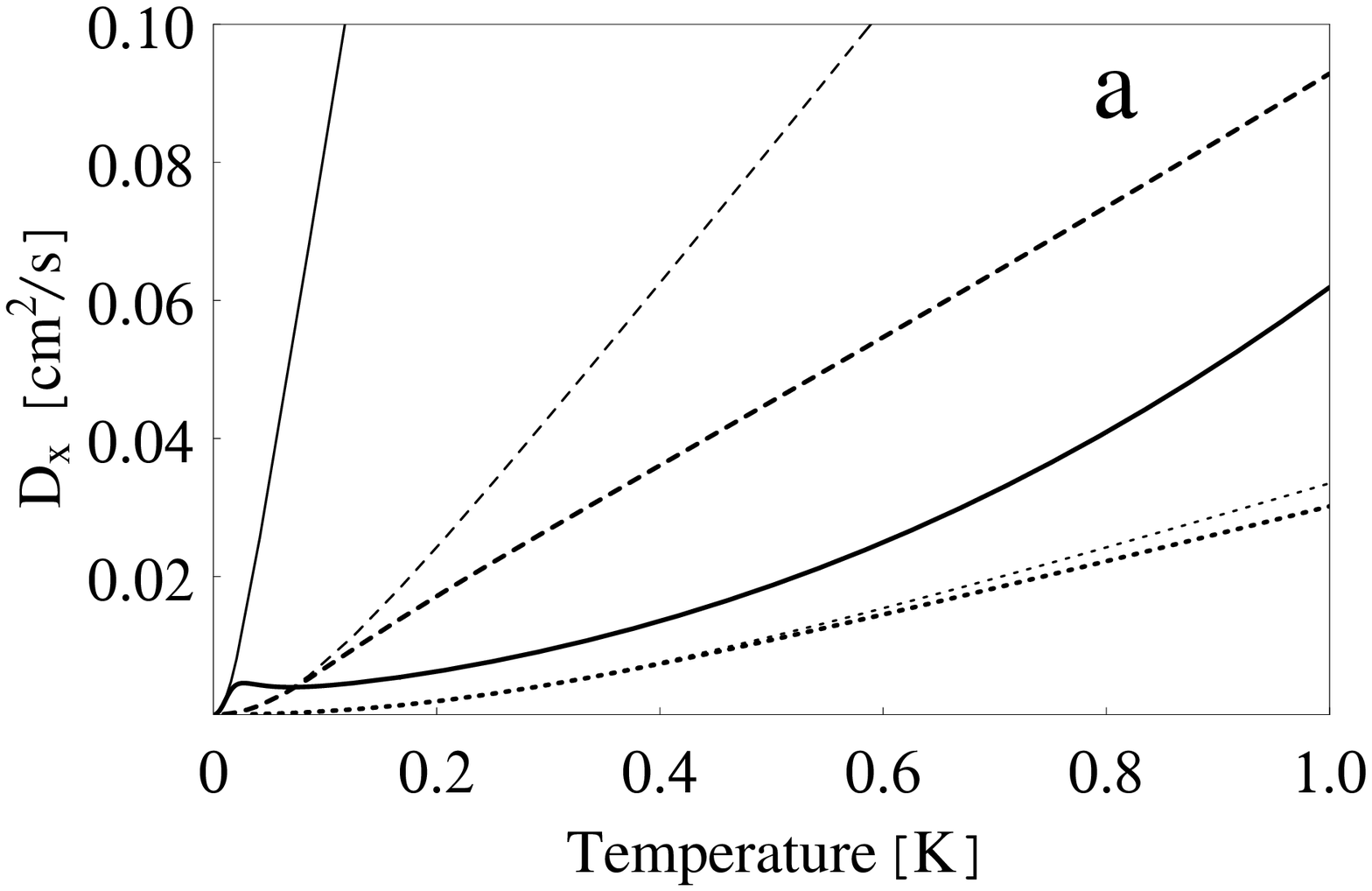,width=3.6in}}
\centerline{\psfig{file=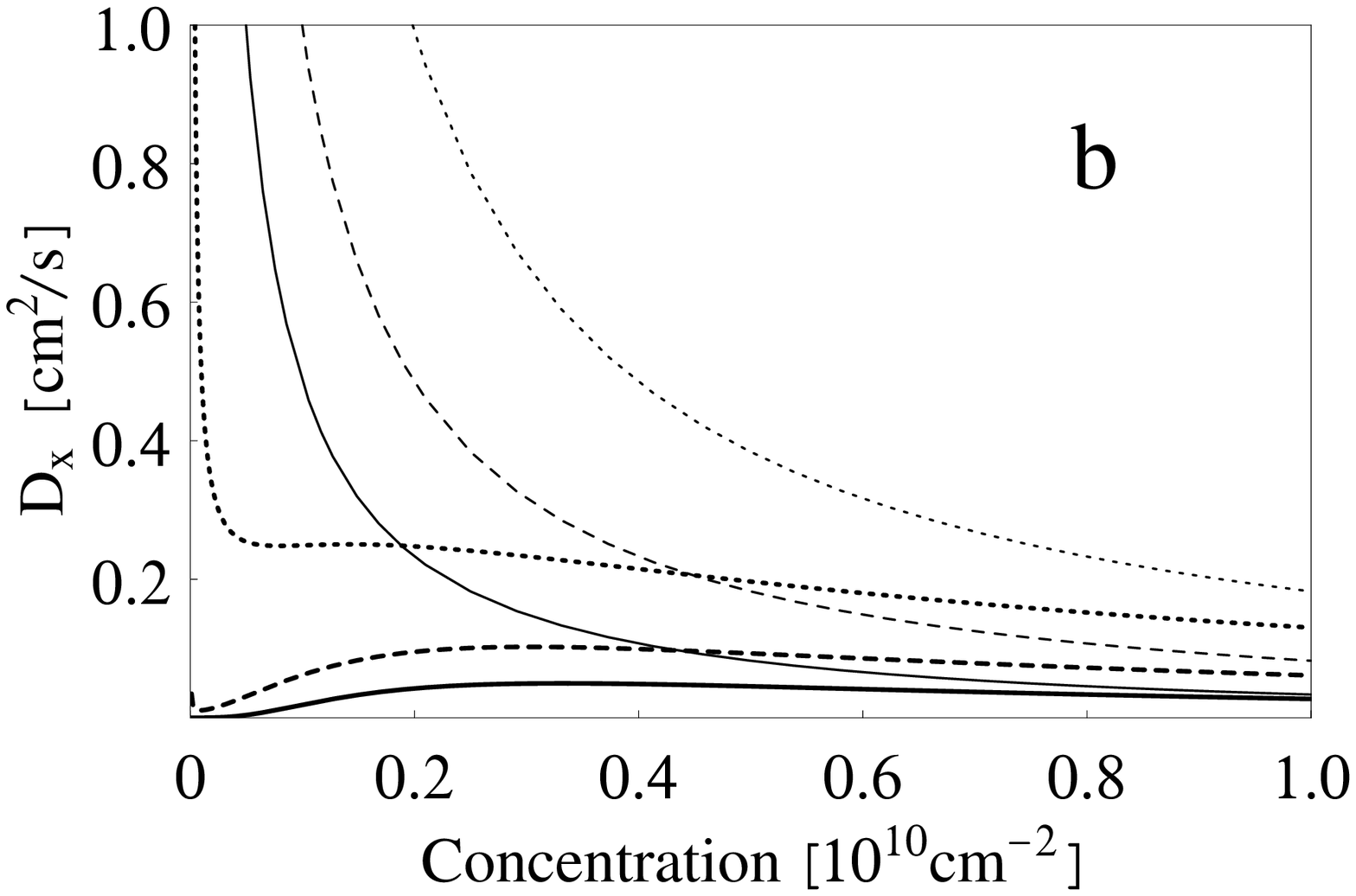,width=3.6in}}
\vspace{0.5cm}
\caption{Diffusion coefficient of indirect excitons in GaAs/AlGaAs 
coupled QWs calculated using Eqs.\,(\ref{diffusion})-(\ref{Ekin}) 
with $C = 4/\pi$ and $\chi = a^{\rm B}_{\rm 2d}/(2d)$. (a) The diffusion 
coefficient versus $T$ for $n_{\rm 2d} = 10^9$\,cm$^{-2}$ (solid lines), 
$5 \times 10^9$\,cm$^{-2}$ (dashed lines) and $2 \times 10^{10}$\,cm$^{-2}$ 
(dotted lines). (b) The diffusion coefficient against $n_{\rm 2d}$ 
for $T = 0.5$\,K (solid lines), 1\,K (dashed lines) and 
2\,K (dotted lines). $D_{\rm x}^{\rm (2d)}$ and 
$\tilde{D}_{\rm x}^{\rm (2d)}$ are shown by the thin and bold lines, 
respectively. $M_{\rm x} = 0.215\,m_0$, $\mu_{\rm x} = 0.046\,m_0$ and 
$d = 11.5$\,nm.}
\end{figure}

\begin{eqnarray}
\label{screen}
&&\delta n_{\rm 2d} = - \, \frac{U_{\rm rand}({\bf r}_{\|}) 
n^{(0)}_{\rm 2d}}{k_{\rm B}T_0(e^{T_0/T} - 1)^{-1} + 
u_0 n^{(0)}_{\rm 2d}} \, , 
\nonumber \\  
&&U_{\rm eff} = u_0 n^{(0)}_{\rm 2d} + 
\frac{U_{\rm rand}({\bf r}_{\|})}{1 + [(2M_{\rm x})/(\pi \hbar^2)] 
(e^{T_0/T} - 1) u_0 } \, , 
\end{eqnarray}
where $\delta n_{\rm 2d} = n_{\rm 2d}({\bf r}_{\|}) - n^{(0)}_{\rm 2d}$.  
$U_{\rm eff} = U_{\rm rand}({\bf r}_{\|}) + u_0  n_{\rm 2d}({\bf r}_{\|})$ is 
the effective, screened in-plane potential. Relaxation of the 
long-range-correlated random potential is described by the denominator in the 
expression for $U_{\rm eff}$ [see Eq.\,(\ref{screen})]. For $n_{\rm 2d}^{(0)} 
\gg |U_{\rm rand}/u_0|$ the denominator is much larger than unity. For 
classical statistics, when $T \gg T_0$, relationships (\ref{screen}) reduce 
to $\delta n_{\rm 2d}({\bf r}_{\|}) = - 
[ U_{\rm rand}({\bf r}_{\|})n_{\rm 2d}^{(0)} ]/[ k_{\rm B} T + 
u_0 n_{\rm 2d}^{(0)} ]$ and $U_{\rm eff}({\bf r}_{\|}) = u_0n^{(0)}_{\rm 2d} 
+ [ U_{\rm rand}({\bf r}_{\|}) k_{\rm B}T ]/[ k_{\rm B} T + 
u_0 n_{\rm 2d}^{(0)} ]$. For $u_0 n_{\rm 2d}^{(0)} \gg k_{\rm B} T$ the 
latter expression describes strong suppression of the potential fluctuations, 
i.e., instead of input $U_{\rm rand}({\bf r}_{\|})$ one gets 
$\kappa U_{\rm rand}({\bf r}_{\|})$ with $\kappa = 
(k_{\rm B}T)/(u_0 n^{\rm (0)}_{\rm 2d}) \ll 1$. The screening effect becomes 
particularly strong in the quantum regime, $T_0 \geq T$. Formally this 
is due to the term $[\exp(T_0/T) - 1]$ in the expressions (\ref{screen}). 

In Fig.\,2 we plot the effective potential $U_{\rm eff}(x) - u_0 
n_{\rm 2d}^{(0)}$ and concentration of excitons, $n_{\rm 2d}(x)$, calculated 
with Eq.\,(\ref{final}) for a model 1D long-range-correlated random potential 
$U_{\rm rand}(x)$, shown by the bold lines, and realistic values of 
$n_{\rm 2d}^{(0)}$, $D_x$, $T$, $\tau_{\rm opt}$ and amplitude of 
$U_{\rm rand}$ (about $0.5$\,meV). Figure\,2 clearly illustrates 
the origin of the screening effect: Accumulation of indirect excitons 
in the minima of $U_{\rm rand}({\bf r}_{\|})$ and their depletion at the 
maxima of $U_{\rm rand}({\bf r}_{\|})$. In the experiments 
\cite{Butov1994,Butov2001,Butov2002} the blue-shift of the energy 
of indirect excitons $\delta U = u_0 n^{\rm (0)}_{\rm 2d} \simeq 1.4 - 
1.6$\,meV for $n^{\rm (0)}_{\rm 2d} = 10^{10}$\,cm$^{-2}$. For this 
moderate concentration of excitons the long-range-correlated in-plane 
disorder is already drastically screened and relaxed at $T \leq 4.2$\,K 
(see also Fig.\,2). One can directly include the disorder-assisted 
effects into diffusion Eq.\,(\ref{final}) by using a thermionic model. 
In this case we replace $D_{\rm x}^{\rm (2d)}$ by 
${\tilde D}_{\rm x}^{\rm (2d)} = D_{\rm x}^{\rm (2d)} \exp\left[ - 
(U_{\rm eff} - u_0 n_{\rm 2d}^{(0)})/ (k_{\rm B}T)\right]$ and remove 
$U_{\rm QW} = U_{\rm rand}$ from eq.~(\ref{final}). For $T \gg T_0$ the 
diffusion coefficient ${\tilde D}_{\rm x}^{\rm (2d)}$ is 
\begin{equation}
\label{thermionic}
{\tilde D}_{\rm x}^{\rm (2d,cl)} = D_{\rm x}^{\rm (2d,cl)} 
\exp \left[ - \frac{U_{\rm rand}({\bf r}_{\|})}{k_{\rm B} T + 
u_0 n_{\rm 2d}^{(0)}} \right] \, , 
\end{equation}
where $D_{\rm x}^{\rm (2d,cl)}$ is given by Eq.\,(\ref{class}) and 
$U_{\rm rand}({\bf r}_{\|})$ can be replaced by $U_{\rm rand}^{(0)} = 2 
<|U_{\rm rand}({\bf r}_{\|})|>$. The calculated diffusion coefficent 
${\tilde D}_{\rm x}^{\rm (2d)}$, which is relevant to the analysis of the 
experimental data \cite{Butov1994,Butov2001,Butov2002}, 
is shown in Fig.\,1. Note that due to the relatively strong 
dipole-dipole interaction only a dip, natural or externally applied, in-plane 
potential $U_{\rm trap} \sim 10$\,meV can spatially confine the indirect 
excitons of high concentrations $n_{\rm 2d}^{(0)} \sim 5 \times 
10^{10}$\,cm$^{-2}$ \cite{Butov2002,Negoita1999}. Quantum diffusion of 
indirect excitons towards a $\mu$m-scale trap and their real-space 
distribution $n_{\rm 2d} = n_{\rm 2d}({\bf r}_{\|})$ in the trap can be 
calculated by using Eq.\,(\ref{final}) with ${\tilde D}_{\rm x}^{\rm (2d)}$ 
and $U_{\rm QW} = U_{\rm trap}({\bf r}_{\|})$ \cite{Butov2002}. 

\section{Discussion}
In the experiments \cite{Butov1994,Butov2001,Butov2002} with 
GaAs/AlGaAs coupled QWs the effective temperature $T$ of indirect excitons 
can considerably exceed the bath, cryostat temperature $T_{\rm b}$. The 
excitons cool down due to interaction with bulk LA-phonons, while incoming, 
optically-generated high-energy excitons tend to increase $T$. Thus in order 
to adapt the quantum diffusion picture to the optical experiments, 
Eq.\,(\ref{final}) should be completed with Eq.\,(11) of \cite{Ivanov1999}. 
The latter describes the local temperature $T=T({\bf r}_{\|},t)$ which is 
$n_{\rm 2d}$-dependent especially for well-developed BE statistics of QW 
excitons. 

The diffusion transport occurs when $l_{\rm 2d}/a_{\rm 2d}^{(0)} 
\gg 1$ and when the exciton-exciton interaction is the dominant 
scattering \ mechanism. The \ above \ conditions limit the

\begin{figure}
\centerline{\psfig{file=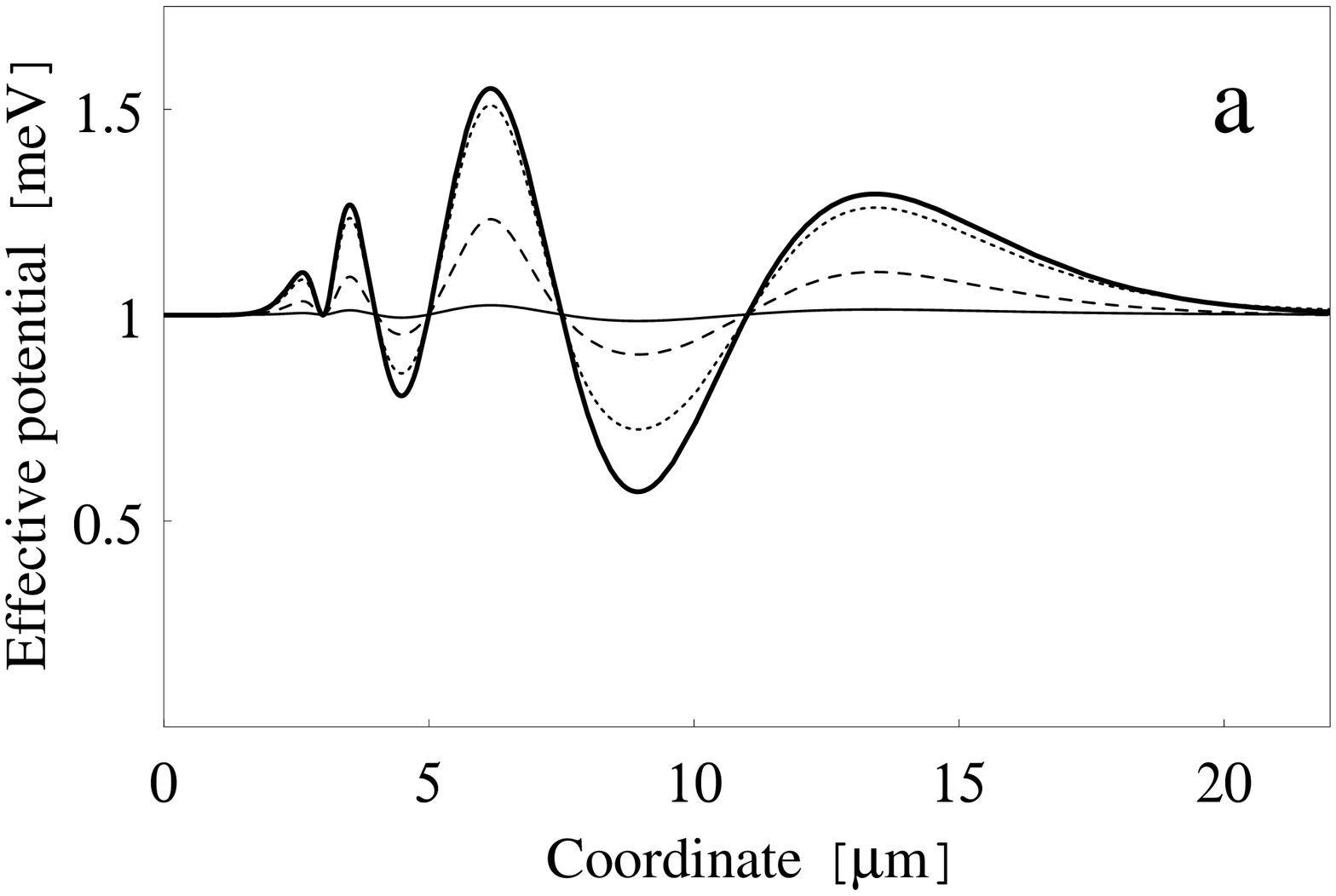,width=3.6in}}
\centerline{\psfig{file=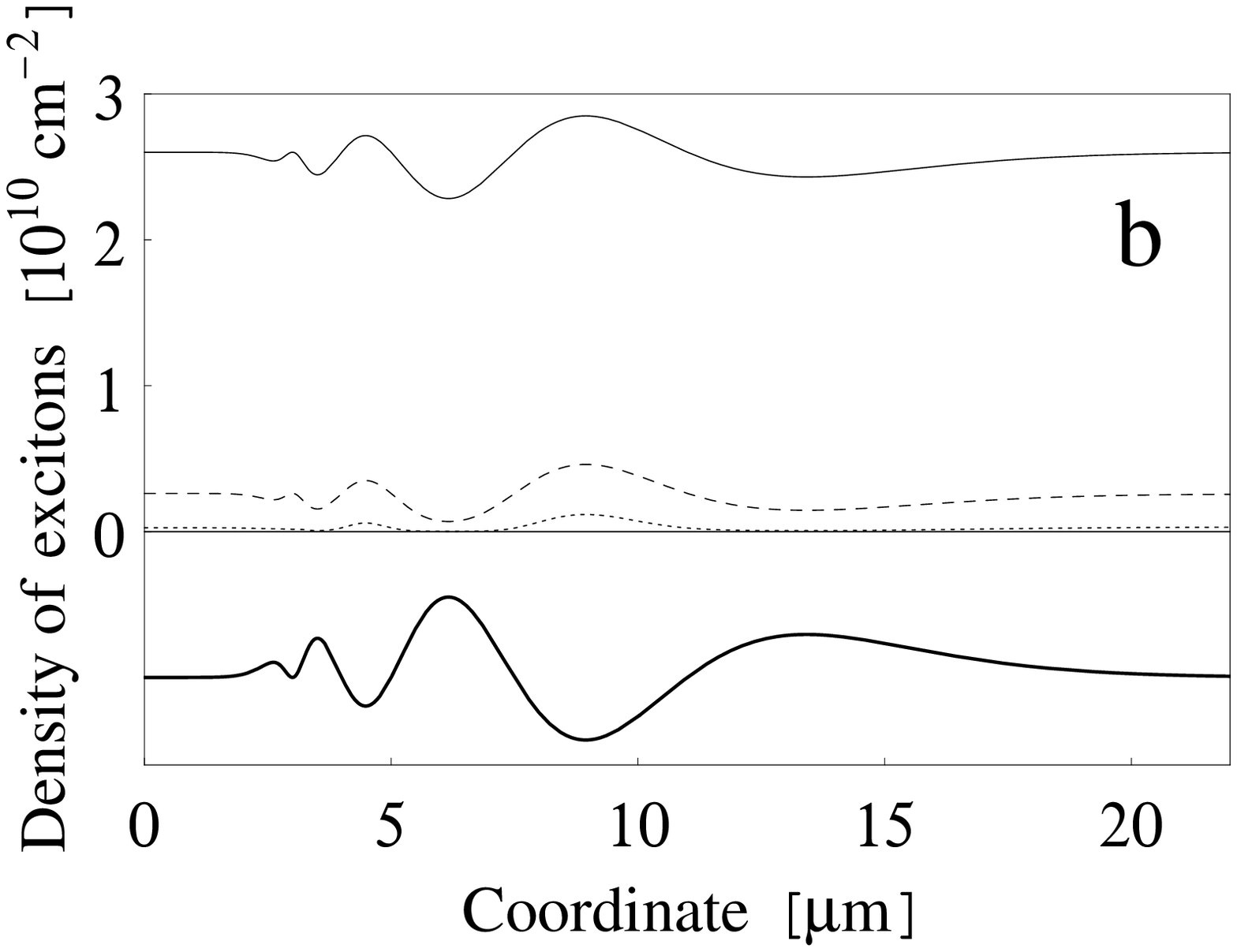,width=3.6in}}
\vspace{0.5cm}
\caption{ Relaxation of long-range-correlated disorder by the dipole-dipole 
interaction of indirect excitons in GaAs/AlGaAs coupled QWs. (a) The 
effective, screened potential $U_{\rm eff}(x) - u_0 n^{(0)}_{\rm 2d}$ 
and (b) local concentrations of indirect excitons $n_{\rm 2d}(x)$ 
versus in-plane coordinate $x$. The average concentrations are 
$n^{(0)}_{\rm 2d} = 2.6 \times 10^{10}$\,cm$^{-2}$ (thin solid line), 
$2.6 \times 10^{9}$\,cm$^{-2}$ (dashed line), and 
$2.6 \times 10^{8}$\,cm$^{-2}$ (dotted line). Temperature $T = 2$\,K, 
diffusion coefficient $D_x=100$\,cm$^2$/s, and radiative lifetime 
$\tau_{\rm opt} = 20$\,ns. In both figures the input, unscreened 
potential $U_{\rm rand}(x)$ is shown by bold solid lines. } 
\end{figure}


\noindent
application of the theory to 
$10^9$\,cm$^{-2} \leq n_{\rm 2d} \leq 2 \times 10^{10}$\,cm$^{-2}$ for 
GaAs-based QWs. The developed model is definitely not applicable for $T \leq 
T_{\rm c}$, where $T_{\rm c}$ is the transition temperature to a superfluid 
phase. Note that for a very dilute gas of quasi-2D excitons $T_{\rm c} \ll 
T_0$ \cite{Popov1972,Fisher1988}. Furthermore, in experiments with 
spatially-inhomogeneous optical generation of QW excitons the diffusive 
propagation always preceds a possible transition to the superfluid motion 
of excitons. 

By analysing a relevant Gross-Pitaevskii equation we have also found the 
screening effect for mid-range-correlated (length scale of a few 
$a_{\rm 2d}^{\rm B}$) in-plane disorder. Namely the dipole-dipole interaction 
between excitons considerably decreases the localization energy of an indirect 
exciton in a mesoscopic in-plane trap. This leads to relaxation of the  
mid-range-correlated QW potential fluctuations. Short-range 2D disorder, which 
can be described in terms of a random contact potential, is also strongly 
suppressed by interacting bosons \cite{Krauth1991}. Thus we conclude that in 
high-quality GaAs/AlGaAs coupled QWs the in-plane random potential of any 
correlation length is effectively screened and practically removed at 
$n_{\rm 2d}^{(0)} \geq 10^{10}$\,cm$^{-2}$, due to the dipole-dipole 
interaction of indirect excitons. In this case the in-plane momentum of 
indirect excitons, ${\bf p}_{\|}$, becomes a good quantum number even for 
low-energy particles, as was demonstrated in the experiments 
\cite{Butov2001,Parlangeli2000}. Thus the phonon-assisted relaxation and 
PL kinetics \cite{Ivanov1999,Ivanov2000,Soroko2002}, formulated in terms 
of well-defined ${\bf p}_{\|}$, are adequate for explanation and modelling 
of the recent experiments on statistically-degenerate excitons in 
GaAs/AlGaAs coupled QWs \cite{Butov2001}.

We attribute a relatively large width $\sim 0.5 - 1$\,meV of the PL line 
associated with indirect excitons at $n_{\rm 2d}^{(0)} \geq 
10^{10}$\,cm$^{-2}$ to intrinsic, homogeneous broadening rather than to a 
disorder-assisted inhomogeneous width. While the dipole-dipole interaction 
of indirect excitons is much weaker and of shorter-range than the $1/r_{\|}$ 
Coulomb law, it gives rise to a relatively large correlation energy in 
comparison with the leading  mean-field-theory correction $\delta U = u_0 
n_{\rm 2d}^{(0)}$. This is because screening of the dipole-dipole interaction 
by indirect excitons is very weak. In this case the optical decay of an 
indirect exciton interacting with its nearest neighbouring exciton(s) gets 
an energy uncertainty. The latter depends upon the distance between the 
interacting particles and results in a large ($\sim 1$\,meV) $T$- and 
$n_{\rm 2d}$-dependent homogeneous width of indirect excitons in GaAs/AlGaAs 
coupled QWs.

\section{Conclusions}
In this work we have developed a quantum diffusion theory for (indirect) 
excitons in (coupled) quantum wells. The main results are (i) the quantum 
diffusion Eq.\,(\ref{final}) with the diffusion coefficient, which is 
calculated within a microscopic picture and is given by 
Eqs.\,(\ref{diffusion})-(\ref{Ekin}) and (\ref{thermionic}), (ii) the 
modified Einstein relation (\ref{mobility}) between the mobility and 
diffusion coefficient of QW excitons, and (iii) effective screening of 
QW disorder by dipole-dipole interacting indirect excitons.

\acknowledgments
The author appreciates valuable discussions with  L. V. Butov, D. S. 
Chemla and L. V. Keldysh. Support of this work by the EPSRC is 
gratefully acknowledged.

\end{document}